\documentclass[12pt,preprint]{aastex}

\shorttitle{Super-FM jets from accretion disks}
\shortauthors{J. Ferreira and F. Casse}

\begin{document}

\title{Stationary accretion disks launching super 
  fast-magnetosonic MHD jets}

\author{Jonathan Ferreira\altaffilmark{1}}
\altaffiltext{1}{Laboratoire d'Astrophysique de
  Grenoble, 414 rue de la piscine, BP 
  53, F-38041 Grenoble, ferreira@obs.ujf-grenoble.fr} 
\and
\author{Fabien Casse\altaffilmark{2}}
\altaffiltext{2}{Institute for Plasma Physics Rijnhuizen,
    P.O. Box 1207, 3430 BE Nieuwegein, Netherlands, fcasse@rijnh.nl}

\begin{abstract}
We present self-similar models of resistive viscous Keplerian disks driving
non-relativistic magnetohydrodynamics (MHD) jets becoming super
fast-magnetosonic. We show that in order to obtain such solutions, the
thermal pressure must be a sizeable 
fraction of the poloidal magnetic pressure at the Alfv\'en surface. 
These steady solutions which undergo a recollimation shock causally
disconnected from the driving engine, account for structures with a high
temperature plasma in the sub-Alfv\'enic region. We suggest that only 
unsteady outflows with typical time-scales of several disk dynamical
time-scales can be produced if the suitable pressure conditions are not
fulfilled.  

\end{abstract}


\section{Introduction}

Self-collimated jets are now commonly observed originating from young
stellar objects (YSOs), active galactic nuclei and galactic binaries
\cite[]{liv97}. All these flows share common properties like being always 
correlated with the accretion phenomenon
\citep{har95,fal96,ser98,gall03}. It has 
long been identified that jet  self-confinement requires the presence of a
large scale magnetic field in order to focus the outflowing plasma
\cite[]{cha80}. The ``universal'' paradigm of jet formation relies on the
occurrence of bipolar magnetic fields threading the accretion disk. As a
consequence, the theory of accretion disks had to be revisited in order to
take into account the mass, angular momentum and energy extractions
achieved by the jet. One notorious modification to the standard picture is
the necessary radial stratification of the disk accretion rate, namely
$\dot M_a \propto r^{\xi}$ ($\xi$ being a measure of the disk ejection
efficiency). For instance $\xi=0$ describes a standard disk with no outflow
while $ 0 < \xi < 1$ stands for an ejecting Keplerian accretion disk
\cite[]{f97}.  If one wishes to obtain the exact ejection efficiency, one
has to solve without any approximation the full magnetohydrodynamics (MHD)
2D structure of the disk.

 Anomalous magnetic diffusivity must be present whithin the disk to allow
 accreting (and rotating) mass to cross  the magnetic field lines whereas
 ejected mass becomes frozen in to the field. Only self-similar solutions
 taking into account the underlying resistive accretion disk hitherto
 provided this description  \citep{fp95,f97,cf00a,cf00b}. Within these
 solutions, once in ideal MHD regime, mass is  magnetically accelerated
 along each field line and must successively cross three MHD critical
 points, namely the slow and fast magnetosonic ones (SM,FM) and the
 Alfv\'enic point (A). So far, none of the self-similar solutions was able
 to obtain both disk and jet flows, the latter crossing the three critical
 surfaces.

Using the same framework, \cite{vla00} solved the ideal  MHD jet equations
and provided new solutions crossing these three critical points. However
since these solutions were not connected to the underlying disk, the issue
of super-FM jet production from accretion disk remained.  In this letter,
we intend 
to present the necessary conditions to get self-similar super-FM jets
(Sect.~2), discuss  the properties of typical solutions (Sect.~3) and
conclude with some astrophysical implications that may be put to the test
of observations.

\section{Role of a sub-Alfv\'enic heating}

Stationary jets are described by a set of axisymmetric ideal MHD
equations. Thus the poloidal magnetic field writes   ${\bf B_p} = (\nabla a
\times {\bf e_{\phi}})/r$, where $a(r,z)=Cst$ describes a surface of
constant magnetic flux.  Disk winds are produced whenever a large scale
magnetic field, close to equipartition with the disk thermal pressure
\cite[]{fp95}, is present over a range in   anchoring radii $r_o$. The
corresponding jet is made of magnetic surfaces nested one around each other
with several integrals of motion. In the non-relativistic case, one gets
($u_p$ poloidal velocity, $\Omega$ angular velocity and $\rho$ density):
(1) the mass to magnetic flux ratio $\eta(a)$ with ${\bf u_p} = \eta
(a){\bf B_p}/\mu_o\rho$; (2) the angular velocity of a magnetic surface
$\Omega_*(a)= \Omega - \eta B_{\phi}/\mu_o\rho r$ and (3) the specific
total angular momentum $L(a)=\Omega_*r^2_A = \Omega r^2 -
rB_{\phi}/\eta$ transported away. Here, $r_A$ is the Alfv\'en radius where
mass reaches the 
Alfv\'en poloidal velocity. In this
letter, we are interested in jets that may be heated by their surroundings
so that an adiabatic description is inadequate. Instead, we will assume
the presence of a heat flux ${\bf q} = \nabla H - \nabla
P/\rho$, where $H$ is the usual enthalpy for a perfect gas. Including this
additional effect, one gets the generalized Bernoulli invariant 
$E(a) + {\cal F}(s,a)  = \frac{u^2}{2} + H + \Phi_G
- r \Omega_* B_{\phi}/\eta$, where ${\cal F}(s,a)= \int^s_{s^+} {\bf q}
\cdot {\bf e_{\parallel}}ds'$  is the heating term that depends on a
curvilinear coordinate $s$ along a given magnetic surface ($s^+$ is roughly
the SM point and ${\bf B_p} = B_p {\bf e_{\parallel}}$). The  total specific
energy provided at the disk surface is $E(a)\simeq \Omega_o^2 r_o^2
(\lambda - 3/2)$ for a thin disk, where $\Omega_o$ is the Keplerian
rotation at the anchoring radius $r_o$ and $\lambda=  L/\Omega_o r_o^2$ is
the magnetic lever arm.  The shape of the magnetic surface is given by the
Grad-Shafranov (GS) equation 
\begin{eqnarray}
 (1-m^2)J_{\phi} &= & J_{\lambda} + J_{\kappa} + J_{\beta}\ ,   \label{eq:Grad} \\
\mbox{where}\ \ \  J_{\lambda} & = &\rho r\left(\frac{d{\cal
E}}{da}+(1-g)\Omega_*r^2\frac{d\Omega_*}{da} +
g\Omega_*\frac{d\Omega_*r^2_A}{da}\right)\nonumber\\ 
J_{\kappa} & = & r\frac{B^2_{\phi}-m^2 B^2_p}{2\mu_o}\frac{d\ln\rho_A}{da} +
m^2\frac{\nabla a}{\mu_or} \nabla\ln\rho \nonumber \\
J_{\beta} &=  & \frac{\rho}{B_p} \left ( \nabla {\cal F} - {\bf q} \right )
\cdot {\bf e_{\perp}}  \nonumber
\end{eqnarray}
\noindent Here, ${\bf e_{\perp}} = \nabla a /|\nabla a|$,
$m=u_p/V_{Ap}$ is the Alfv\'enic Mach number and  
$g= 1 - \Omega/\Omega_*$. GS equation provides $a(r,z)$ for a given set of
invariants. Unfortunately, it is a PDE of mixed type: it is
hyperbolic between the cusp (where $u_p=V_c = C_s V_{Ap}/\sqrt{C_s^2 +
  V_A^2}$) and the slow-magnetosonic surface (where 
$u_p=V_{SM,p}$), elliptic between the SM and the fast-magnetosonic surface
(where $u_p= V_{FM,p}$) and hyperbolic further out. The magnetosonic
phase speeds involved in these definitions are the usual ones, i.e.
waves travelling along the poloidal field.
Solving Eq.~(\ref{eq:Grad}) remains a major challenge in applied
mathematics: it would require to {\em a priori} know the locus of 
these surfaces, whereas they emerge as the global solution evolves. In
practice, one either solves the time-dependent problem with full MHD codes,
or uses a method of variable separation.

\begin{figure}[t]
  \plotone{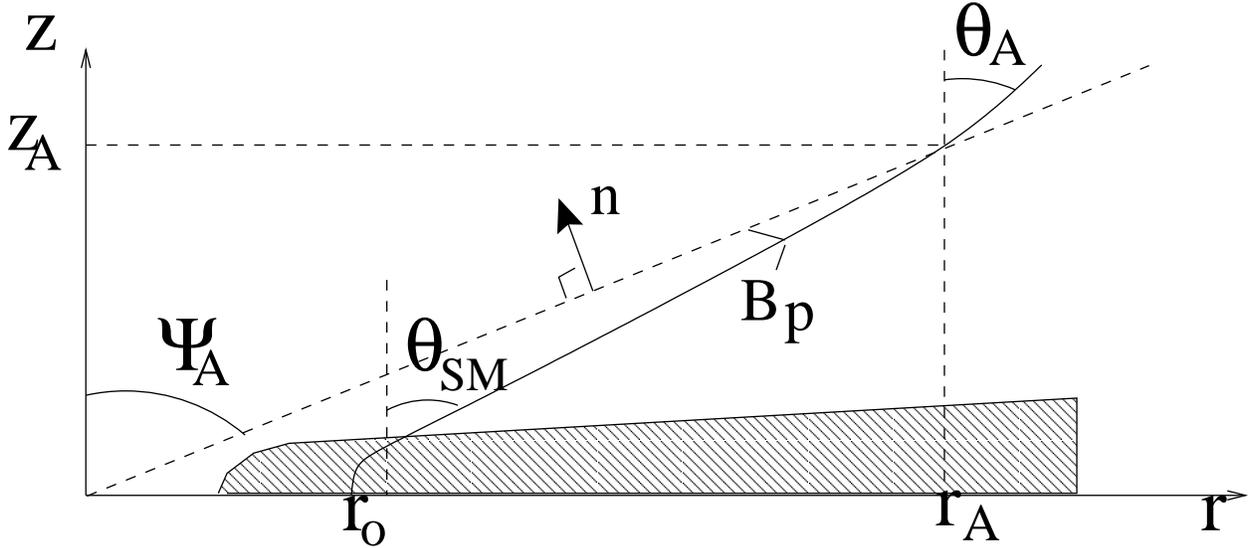}
   \caption{Geometry of a disk wind configuration and definitions of
   quantities related to the Alfv\'en critical surface. The unit vector
   ${\bf n}$ is defined as ${\bf (z,r)}/\sqrt{r^2+z^2}$. }
   \label{fig:ang}
\end{figure}

Self-similar solutions allow to solve the full set of MHD equations without
any approximation. The problem reduces to propagate the solution along a
self-similar variable $x=z/r$, which involves the inversion of a
matrix. Its determinant vanishes at three singular points where the
following numbers become equal to unity: $M_{SM} = V/V_{SM,n}$, $M_A=
V/V_{A,n}$ and $M_{FM} = V/V_{FM,n}$, where $V= {\bf u_p}\cdot {\bf n}$ and
$V_{A,n}= {\bf V_A} \cdot {\bf n}$ are projections in the direction $\bf n$
(see  Fig.~\ref{fig:ang}) and $V^2_{SM/FM,n} = \frac{1}{2} \left( V_A^2 +
C_s^2 \pm \sqrt{ (V_A^2 + C_s^2)^2 - 4C_s^2V^2_{A,n}}\right)$
\cite[]{fp95}. Not all these critical points    coincide with the above
mentionned points where the flow changes type: $M_A= m$, $M_{SM} \simeq
u_p/V_c \simeq u_p/V_{SM,p}$ (in the cold Keplerian limit)    but $M_{FM} <
n= 
u_p/V_{FM,p}$. The 
critical FM surface is always located downstream in the hyperbolic region.
The necessary conditions to provide cold super-Alfv\'enic jets from
Keplerian disks were given in \cite{f97}.  Once fulfilled, a super-A
solution propagates further away until collimation by the hoop-stress takes
place.  This produces an unavoidable decrease of the projected velocity ($V
\rightarrow -u_{r} \simeq 0$), even if the poloidal velocity reaches its
asymptotic value of $\Omega_{o}r_{o} \sqrt{2\lambda -3}$. The only way to
allow for a super-FM solution with $M_{FM}>1$ is to lower this projection
effect by forcing the magnetic surfaces to remain wide open, namely
bringing the Alfv\'en surface closer to the disk (increase $\Psi_{A}$).
The self-similar expression of the GS equation in the {\em cold} regime can
be written as $\cos \theta_A = R(\theta_A;\Psi_A)$ at the Alfv\'en point
\cite[]{cf00a}. It is an implicit equation  providing the jet opening angle
$\theta_A$ for a given position of the Alfv\'en surface $\Psi_A$. In the
cold limit, one has $\tan \theta_{SM} \sim \tan \Psi_A (1- \lambda^{-1/2})$
for a given magnetic lever arm $\lambda \simeq r_A^2/r_o^2$ and initial
opening angle $\theta_{SM}$ (see Fig.~\ref{fig:ang}). This initial opening
angle is constrained by the underlying disk vertical equilibrium. The
larger angle, the larger the magnetic compression and the less mass is
being ejected. Only angles up to $\sim$ 45\degr\ ($z_A \sim r_A$) have been
proved to be possible from Keplerian accretion disks, either with
isothermal \cite[]{f97} or adiabatic jets \cite[]{cf00a}, but none of these
solutions can become super-FM.

Every super-FM solutions obtained by \cite{vla00} exhibit Alfv\'en surfaces
closer to the equatorial plane (i.e. $\Psi_A \sim$ 60\degr ), but being not
connected to a resistive MHD disk, they did
not have to fulfill the requirement of a quasi-static disk vertical
equilibrium. Actually no Keplerian disk would probably survive the
overwhelming magnetic compression imposed by the bending of the field lines
(or allow the imposed mass effluvium). The only possibility to have an
Alfv\'en surface closer to the disk is to break this univocal link between
$\Psi_{A}$ and $\theta_{SM}$. This implies a change of some invariants
(entropy and total specific energy) between the disk and the Alfv\'en
surface. Physically this requires an extra force term in the GS equation,
namely a strong outwardly directed pressure gradient in the sub-A
region. Within the self-similar framework, it means building up a large
thermal pressure, thus an additional heating starting above the disk.
\begin{figure}[t]
  \plotone{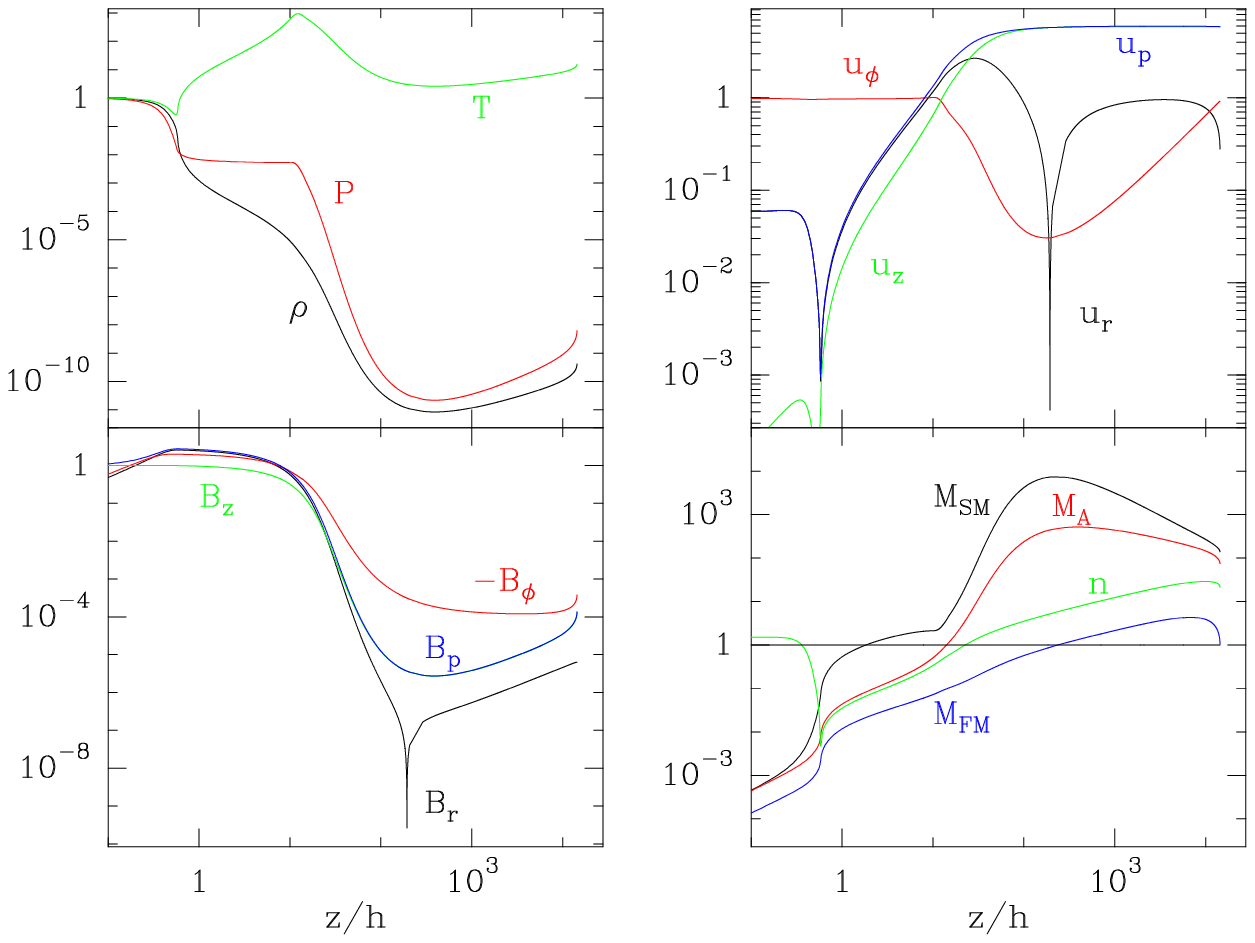}
  \caption{Typical super-FM disk wind with $\xi=0.03, \epsilon=
     0.03$ ($h = \epsilon r$). Density, pressure and temperature are
     normalized to their value at the 
     disk midplane, the magnetic field components to $B_z(z=0)$ and 
     the velocities to the Keplerian speed at the anchoring radius
     $r_o$. All magnetic field components remain comparable from
     the disk surface to the Alfv\'en point. Note that the density profile
     inside the disk, where both $u_r$ and $u_z$ are negative, is very
     different from a gaussian. Recollimation takes place at $z\simeq 3\ 10^3 r_o$.}  
   \label{fig:sol}
\end{figure}
The generalized GS equation becomes $\cos \theta_A =
R(\theta_A;\Psi_A) + R_{\beta} (\theta_A;\Psi_A)$ where
\begin{eqnarray}
  R_{\beta}&(\theta_A;\Psi_A)& = -\ \frac{g_A \beta_A}{4} \left ( 
    \frac{2}{g_A} \cos \theta_A 
    + \frac{\sin \Psi_A}{\sin(\Psi_A - \theta_A)} \right .\label{eq:GSth}\\
   &\times& \left \{ \frac{ {\cal  F}_A} {C^2_{s,A}} - \frac{2}{g_A} - 
      \frac{d \ln \rho_A}{d \ln r_o} - \frac{1}{\gamma -1} \right \}
    \nonumber  \\
    & +& \left . \frac{\cos (\Psi_A -\theta_A)}{\sin \Psi_A}  \left \{ 
        \frac{r_A {\bf q \cdot e_{\parallel}}_A }{C^2_{s,A}\cos \theta_A}
        - \left . \frac{\partial \ln C^2_s}{\partial x} \right |_A 
      \right \}
    \right )\nonumber
\end{eqnarray}
is the contribution of this additional heat flux and $\beta_A$ is the ratio
of the plasma pressure $P_A \equiv \rho_A C^2_{s,A}$ to the poloidal
magnetic pressure at the Alfv\'en point. This equation shows that $\beta_A$
large enough ($\beta_A \la 1$) and $R_{\beta}$ negative are two necessary
conditions to increase $\Psi_A$. Indeed, since $\theta_A$ is always smaller
than $\Psi_A$ , any tendency to increase $\theta_A$ leads to a lowering of
the Alfv\'en surface.
 
At the Alfv\'en surface, $\beta_A = 2 \omega_A^2 \frac{\epsilon^2}{\lambda}
\frac{T_A}{T_o}$ where $\omega_A = \Omega_* r_A/u_{p,A} \ga 1$
\citep{f97,cf00a} and $\epsilon= h/r$ is the disk aspect ratio. This
general expression shows that any cold jet (isothermal $T_A = T_o$ or
adiabatic $T_A \ll T_o$) always displays $\beta_A \ll 1$. In order to have
any influence on the transverse equilibrium, this additional heating must
provide a large increase in jet temperature, namely $T_A \ga
T_o/\epsilon^2$.     The second condition ($R_{\beta} <0$) sheds light on
the required heating function. The sum of the first two terms in
(\ref{eq:GSth}) is usually always negative. Indeed, energy conservation
gives ${\cal F}_A/C^2_{s,A} \geq \frac{\gamma}{\gamma-1} (1 -
\frac{T_o}{T_A})$ because of the tremendous cooling due to the jet
expansion. Since $T_A$ must be large, the ratio ${\cal F}_A/C^2_{s,A}$ is
always large enough (but of order unity). The third term of (\ref{eq:GSth})
shows that the most favourable situation is the presence of additional
heating mainly in the sub-A region, i.e. a vanishing heat flux (${\bf q
\cdot e_{\parallel}}_A =0$ or very small) and an already decreasing
temperature (due to adiabatic cooling).

\section{Self-similar numerical solutions}

We follow basically the same integration procedure as in our previous works
(see \cite{cf00b} for more details). A heating function is assumed to be
present, starting at the disk surface but vanishing before the Alfv\'en
point, with an 
adiabatic index $\gamma=5/3$. Further up, we allow for a continuous
transition to a polytropic energy equation, $P \propto \rho^{\Gamma}$. The
X-type FM critical point allows to determine the critical value $\Gamma_c$
of the polytropic index: 
if $\Gamma < \Gamma_c$ thermal acceleration is too inefficient
(breeze-like solution) whereas if $\Gamma > \Gamma_c$ the strong decrease in
enthalpy leads to a shock-like solution. Although \cite{vla00} used an
analogous way, our solutions strongly differ by the fact that jet
invariants are fixed by the disk. Therefore, we need first to drastically
increase the jet enthalpy before fine-tunning the polytropic index. As in
solar wind models, we are playing around with one free parameter ($\Gamma$)
whilst one should solve the full energy equation.
 
Figure~\ref{fig:sol} shows a typical super-FM solution obtained with
$\Gamma_c = 1.45$. The energy input required can be measured by the
ratio $f= {\cal  F}(x_A,a)/E(a)$ since most of the heating occurs in the
sub-A region. Solutions displayed here required $f$ of several $10^{-3}$,
allowing to get $\Psi_A \simeq$ 65\degr\  with $\beta_A \simeq 0.1$ (condition $T_A \sim T_o/\epsilon^2$ is
verified). Note that smaller temperature values would also allow super-FM jets but they
would just be terminated much sooner as in \cite{vla00}.   
\begin{figure}
\plotone{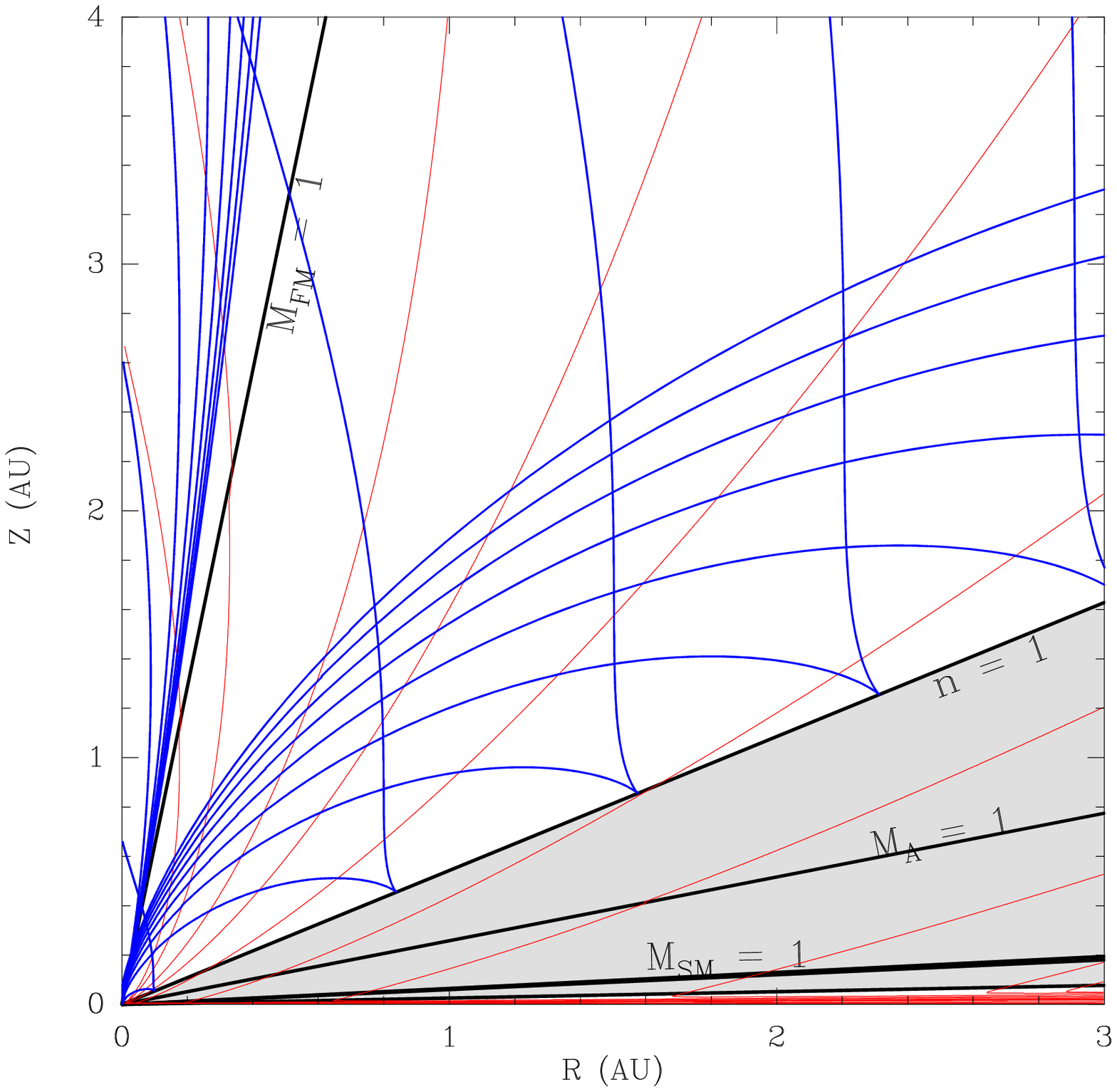}
   \caption{Poloidal cross section showing the three critical surfaces (SM,
     A and FM), some characteristics (blue lines, in the hyperbolic
     domain) as well as the two elliptic regions (shaded). Contrary to
    \citet{vla00}, the streamlines (red lines) are computed from the
     midplane of the resistive accretion disk.  This solution has
     $\xi=0.09, \epsilon= 0.03$ and $\Gamma_c= 1.56$.} 
   \label{fig:cara}
\end{figure}
\noindent In general 2D flows, the ``causal horizon'' (here the $M_{FM}=1$
surface) 
is the envelope of one of the two families of characteristics
\cite[]{tsin96} and not the surface of parabolicity $n= 1$ (see
Fig.~\ref{fig:cara}). Any
perturbation occurring to the flow downstream to the $M_{FM}=1$ surface is
unable to cross this horizon. This result is generic to 2D solutions, the
only bias introduced by self-similarity is the conical shape of such
surfaces, not their separate existence. This has strong consequences on
numerical experiments, as already pointed out by \cite{ust99}. To ensure
the absence of feedback from the imposed boundary conditions, the Mach
cones (defined locally as the tangents to the characteristics) must be
directed out of the computational domain at its boundaries. 
 
\section{Astrophysical implications}

The present computed MHD flows are the first-ever steady-state solutions
describing an overall accretion-ejection structure from the resistive
accretion disk to the super-FM jet region.  The strict stationarity of such
accretion-ejection engines depends critically on the thermal properties of
the sub-Alfv\'enic region. If the plasma pressure, measured at the Alfv\'en
point, is a sizeable  fraction of the poloidal magnetic pressure, MHD jets
from Keplerian accretion disks can become super-FM. In the super-FM region,
the jet is always facing a recollimation that ends up as a shock.  The
further jet propagation  requires numerical time-dependent
simulations. Around a protostar, such thermal pressure gradient occurs
whenever temperatures as high as several $10^5$ K are reached along the
inner streamline. This is compatible with recent observations of
blueshifted UV emission lines \cite[]{gom01} and some absorption features
\cite[]{tak02}. Unfortunately, the heating source can only be inferred from
its effects and its origin remains a crucial issue. For instance YSOs
accretion disks are assumed to be highly magnetized so one may safely
expect that some accretion energy is also dissipated in the upper disk
layers and  provides coronal heating \citep{gal79,hey89}. This is actually
shown by both numerical simulations \cite[]{mil00} and some observational
indication of accretion powered coronae \cite[]{kwan97}. Moreover, since
the central object has a hard surface, the shock of the infalling material
provides another source of UV radiation (as well as X-rays), illuminating
the disk and heating the sub-A region \cite[]{fer03}. An alternative to
this additional heating would be the presence of a high pressure inner flow (a
``spine'') forcing the MHD disk wind to
open up. In YSOs such a flow could be provided by the
interaction between the protostellar magnetosphere and the disk
\citep{fpa00,mat02,rom02}. Temperatures required around a compact object
imply a relativistic plasma. In this case, the inner pressure could be
provided by an inner beam composed of relativistic electron-positron pairs,
heated and accelerated inside the hollow part of the disk wind
\cite[]{ren98}.

On the other hand if an accretion-ejection engine
cannot provide this additional heating or if there is no inner ``spine'',
then the thermal pressure is negligible at the Alfv\'en surface and jets
remain sub-FM. Recollimation towards the axis leads to the formation of a
shock and the overall structure is therefore unsteady. However, no MHD
signal  can propagate upstream along the magnetic field towards the
disk. Instead, in this hyperbolic region, the information that a shock has
occurred is first carried away by MHD waves travelling along the
characteristics until the $n=1$ surface is reached (see
Fig.~\ref{fig:cara}). Once in the elliptic domain, the fastest mode travels
along the magnetic field down to the disk. As a  consequence, the time
taken by MHD waves to inform the disk is always larger than the one given by
e.g. computing the time $\tau = \int ds/V_{FM,p}$ taken by the fast mode
along the same fieldline.  As an illustration, let us take the cold sub-FM
solutions of \cite{f97}, as pictured in his Fig.~6. For the solutions
recollimating right  after the Alfv\'en surface this time $\tau$ is roughly
equal to the orbital period $\tau_o$ at the anchoring radius, whereas for
those recollimating much farther away, one gets $\tau \ga 10^2 \tau_o$.
The presence of sporadic jet events with time scales much larger than disk
dynamical time scales (\cite{rag02}, \cite{gall03} and references therein)
could fit into the picture of an accretion-ejection engine trying to adjust
itself.    

The amount of large scale poloidal magnetic flux trapped in accretion disks
is completly unknown. The above astrophysical implications only hold if this
flux is large enough so that an equipartition field spans at least one
decade in radius in the disk. Indeed, in such circumstances, there is no
physical reason for strong gradients in jets and one may expect an almost
plane Alfv\'en surface \citep{kra99,cas04}. This kind of jets display
dynamical properties (acceleration and collimation) that weakly depend 
 on the radial (inner and outer) boundary conditions, as in
self-similar models. On the contrary, if the flux is small and
concentrated at the inner edge of the disk, one would expect an almost
spherical expansion of the field lines as in X-wind models \cite[]{shu94}.
Observations allowing to infer jet velocity patterns and to relate them to the
source \citep{gar01,bac02,pes03} are necessary to discriminate between
these two extreme pictures.
\acknowledgements
 F.C. is a postdoctoral fellowship of the European Community's Human
Potential Programme  PLATON under contract HPRN-CT-2000-00153. F.C. would
also like to thanks the team SHERPAS for its hospitality during his stay at
the LAOG.

\end{document}